\documentclass{aa}  
\usepackage{graphicx}
\usepackage[export]{adjustbox}
\usepackage{subcaption}
\usepackage[labelfont=bf]{caption}

\newcommand{\uproman}[1]{\uppercase\expandafter{\romannumeral#1}}
\usepackage{txfonts}
\usepackage{color}
\usepackage{hyperref}
\begin{document}

\title{NLTE modelling of integrated light spectra}

\subtitle{Abundances of barium, magnesium, and manganese in a metal-poor globular cluster}

\author{P. Eitner\inst{1,2} \and M. Bergemann\inst{2} \and S. Larsen\inst{3}  }

\institute{Ruprecht-Karls-Universit\"at, Grabengasse 1, 69117 Heidelberg, Germany\\
\email{eitner@mpia-hd.de}
\and
Max-Planck-Institute for Astronomy, K\"onigstuhl 17, 69117 Heidelberg, Germany\\
\email{bergemann@mpia-hd.de}
\and  
Department of Astrophysics/IMAPP, Radboud University, Heyendaalseweg 135, 6525 AJ Nijmegen, The Netherlands\\
\email{S.Larsen@astro.ru.nl} 
}

% \abstract{}{}{}{}{} 
% 5 {} token are mandatory
 
\abstract
{}
{
We study the effects of non-local thermodynamic equilibrium (NLTE) on the abundance analysis of barium, magnesium, and manganese from integrated light spectroscopy, as typically applied to the analysis of extra-galactic star clusters and galaxies. In this paper, our reference object is a synthetic simple stellar population (SSP) representing a mono-metallic $\alpha$-enhanced globular cluster with the metallicity [Fe/H]$=-2.0$ and the age of 11 Gyr.
}
{We used the MULTI2.3 program to compute LTE and NLTE equivalent widths of spectral lines of Mg I, Mn I, and Ba II ions, which are commonly used in abundance analyses of extra-galactic stellar populations. We used ATLAS12 model atmospheres for stellar parameters sampled from a model isochrone to represent individual stars in the model SSP. The NLTE and LTE equivalent widths calculated for the individual stars were combined to calculate the SSP NLTE corrections.}
{We find that the NLTE abundance corrections for the integrated light spectra of the the metal-poor globular cluster are significant in many cases, and often exceed 0.1 dex. In particular, LTE abundances of Mn are consistently under-estimated by $0.3$ dex for all optical lines of Mn I studied in this work. On the other hand, Ba II, and Mg I lines show a strong differential effect: the NLTE abundance corrections for the individual stars and integrated light spectra are close to zero for the low-excitation lines, but they amount to $-0.15$ dex for the strong high-excitation lines. Our results emphasise the need to take NLTE effects into account in the analysis of spectra of individual stars and integrated light spectra of stellar populations.}
{}
\maketitle
%
%________________________________________________________________
%
\section{Introduction}
The chemical composition of stellar populations provides vital constraints on their formation histories and on the relative contributions of different nucleosynthetic processes to chemical enrichment. Measurements of overall metallicities and metallicity gradients can provide information on the importance of in- and outflows, variability of initial mass function and star formation activity, and mergers \citep[e.g.][] {Pagel1975,Hartwick1976, prantzos2000, Dalcanton2006, koppen2007, sanchez2009, Kirby2010, lian2018}. 

However, a far more comprehensive picture can be obtained from measurements of individual element abundance ratios. The ratio of alpha-capture elements (O, Mg, Si, S, Ca) to iron can constrain the relative importance of core-collapse and Type Ia SNe, and hence the time scales for chemical enrichment \citep[e.g.][]{Tinsley1979,McWilliam1997,Weinberg2018}. A prime example is the super-solar [$\alpha$/Fe] ratios in massive early-type galaxies, which suggests that the bulk of the stars there formed in intense bursts at intermediate to high redshifts (e.g. \citealt{Thomas2005a,Thomas2010,Maiolino2018}). Heavy elements (e.g. Y, Ba, La, Eu) are sensitive to the rapid and slow neutron capture processes. For example, the enhancement in the [Ba/Fe] ratio in metal-rich stars in the Large Magellanic Cloud may indicate an important contribution by the s-process via AGB winds \citep{VanderSwaelmen2013,Nidever2019}.\\

The chemical abundance patterns in globular clusters (GCs) show additional complexity that cannot be accounted for by standard models for chemical evolution. In particular, the abundances of many of the light elements deviate from those found in the field, with some elements being enhanced (He, N, Na, Al) and others depleted (C, O, Mg) compared to the composition seen in field stars (cf. \citealt{Bastian2018} and references therein). The underlying cause of these abundance anomalies remains poorly understood, although it is generally believed to be related to proton-capture nucleosynthesis at high temperatures  \citep[e.g.][]{renzini2015}.\\

Accurate abundance analysis requires physically realistic methods to compute radiative transfer. It was shown \citep[e.g.][]{Asplund2005a} that one of the main reasons for systematic uncertainties is the assumption of local thermodynamic equilibrium (LTE) instead of the more general approach of non-LTE (NLTE, or kinetic equilibrium after \citealt{Hubeny2014}). NLTE models provide a more realistic way of treating the radiation transfer in the stellar atmosphere. Detailed studies have been performed for 
H \citep[e.g.][]{Barklem2000,Amarsi2016}, 
Li \citep[e.g.][]{Asplund2003,Lind2009}, 
C \citep[e.g.][]{Fabbian2006,Amarsi2018}, 
N \citep[e.g.][]{Caffau2009}, 
O \citep[e.g.][]{Asplund2004,Amarsi2015}, 
Si \citep[e.g.][]{bergemann2013,Zhang2016,Shchukina2017}, 
Ca \citep[e.g.][]{Mashonkina2017},
Na \citep{Baumueller1998,gehren2004},
Mg \citep[e.g.][]{Osorio2015a, Bergemann2017, alexeeva2018}, 
Al \citep[e.g.][]{baumueller1997, Nordlander2017},
S \citep{takeda2005},
Cr \citep[e.g.][]{Bergemann2010},
Ti \citep[e.g.][]{Bergemann2011}, 
Mn \citep[e.g.][]{Bergemann2007, Bergemann2008}, 
Fe \citep[e.g.][]{korn2003, Bergemann2012Fe, Lind2017}, 
Co \citep{bergemann2010b}, 
Zn \citep{takeda2005},
Cu \citep{yan2015,korotin2018},
Ba \citep[e.g.][]{Mashonkina1996,Korotin2010},
Sr \citep[e.g.][]{Short2006, bergemann2012Sr},
and Eu \citep[e.g.][]{Mashonkina2000}.
NLTE effects in the spectral lines and in continua have various origin. Broadly, they can be attributed to the effects of radiation field on the opacity and on the source function. These are, in turn, caused by the influence of radiative processes on the excitation and ionisation equilibria of the element. Both processes may act in a very different manner for different ionisation stages of the same element and in different stellar conditions, being typically more extreme in metal-poor and hotter atmospheres, but also at lower $\log g$, owing to harder UV radiation fields and lesser efficiency of collisions to thermalise the system. We refer the reader to \cite{Bergemann2014a} for a review on the subject.

NLTE effects on the properties of model atmospheres of cool stars have been a subject of several studies. \citet{Short2005} explored the effects of NLTE on the structure of the solar model. Their analysis suggests that the effects on the temperature structure of the models are very small, and do not exceed 20-40 K in the outer layers, with the largest difference of about 100 K deep in the sub-photospheric layers. We note, though, that in those layers the effects of convection are more significant. Also \citet{Haberreiter2008} find that pronounced NLTE effects are manifest in the UV, where NLTE over-ionisation leads to the continuum brightening, and has non-negligible effects on the line opacity. For the models of cool RGB stars, \citet{Young2014} find differences between 1D LTE and 1D NLTE of the order 25 to 50 K only, although the effect depends on the choice of the observable quantity.\\

Very recently, several studies took the effects of NLTE into account in modelling integrated light spectra. \citet{young2017,young2019} considered the effects of NLTE and CMD discretisation for a set of populations with a global metallicity [M$/$H] of $-1.49$, $-1.$, and $-0.55$. Their analysis emphasised the effect of NLTE on photometric colour indices, but they also find non-negligible difference in the lines of Fe, Ti, and O. Also \cite{Conroy2018} considered the effects of NLTE in Na and Ca in modelling the spectra of simple stellar populations on the basis of three model atmospheres. They find small effects, however, their analysis was limited to a SSP with a solar metallicity, where the NLTE effects for these two elements are typically known to be rather small \citep{Asplund2005a}.\\

In what follows we will discuss the NLTE effects on the chemical abundance analysis of integrated light spectra. We deliberately limit ourselves to the analysis of one simple reference synthetic stellar population and three chemical elements, each with a limited number of investigated transitions. Our SSP  represents an 11 Gyr mono-metallic globular cluster with a metallicity of [Fe/H] $= -2$ and a standard $\alpha$-enhancement of [$\alpha$/Fe]$=0.4$ dex, following the work of \cite{Larsen2012,Larsen2014,Larsen2018}. The $\alpha$-enhancement has a typical value found in GCs in the Milky Way, for example, NGC 4372 \citep{Roman2015} and 4833 \citep{Carretta2014}. As noted above, most GCs are known to host multiple stellar populations \citep[e.g][]{carretta2009a,carretta2009b,piotto2015, marino2017,milone2018}, that may add to complexity in the analysis of elemental abundances. While a detailed modelling of these effects is beyond the scope of this paper, quantifying their impact on integrated-light observations requires an adequate treatment of other relevant effects, such as the NLTE corrections considered here. Furthermore we employ new model atoms for Mg, Ba, and Mn, and discuss the NLTE effects for individual stars and for the combined stellar population. \\

This paper is structured as follows. In section \ref{s_methods}, we discuss the model atoms, stellar atmospheres, and the determination of NLTE abundance corrections for the synthetic stellar population. Section \ref{s_results} presents the NLTE corrections for the individual stars and for the integrated light spectra. We close with a short discussion of the results and outline the perspectives.
\section{Methods} \label{s_methods}
The main quantity needed in abundance analysis is the outgoing flux for the model of a stellar atmosphere, which is usually defined by four parameters $T_{\rm eff}$, $\log g$, chemical composition, and, in the case of 1D analysis with hydrostatic models, micro-turbulence. With this flux one derives the equivalent width (EW) of the spectral lines. The comparison of the observed and model EWs yields the desired abundances \citep{Norris2017}. \\
Furthermore, it is common to compute NLTE abundance corrections for stellar atmosphere models by comparing the grids of NLTE and LTE EWs. MULTI\footnote{The publicly available version MULTI 2.3, 2015, \url{http://folk.uio.no/matsc/mul23}} \citep{Carlsson1986} is the most commonly used code, which is well suited for such calculations. MULTI includes a standard handling of background opacity, including line and continuum opacity. The background line opacities are taken into account by reading an opacity sampling file, which was computed with an up-to-date atomic and molecular linelist from \citet{Bergemann2012c} and \citet{Bergemann2015} but excluding the elements treated in NLTE. In this paper, we use the 1-dimensional (1D) version of MULTI with the local version of the lambda operator. A 3D version of MULTI has been presented by \cite{Leenaarts2009}. However, full 3D NLTE calculations are computationally very expensive and cannot be easily applied in the quantitative assessment of NLTE effects in an arbitrarily SSP model. We hence restrict this study to a 1D analysis of NLTE effects and reserve 3D NLTE calculations for future work.

In the following, we briefly describe the NLTE model atoms and the atmospheric models. 
%
%
% Table lines
%
\begin{table}
\begin{center}
\caption{Ba, Mg and Mn lines of interest and their oscillator strength $\log_{10}(gf)$.}
\label{t_lines}
\setlength{\tabcolsep}{0.03\linewidth}
\begin{tabular}{l c c}
\hline
\noalign{\smallskip}
Element/Ion & Wavelength [$\AA$] & $\log_{10}(gf)$ [dex]  \\ 
\noalign{\smallskip}
\hline
\noalign{\smallskip}
\textbf{Barium \uproman{2}}             & 4554.03                        & 0.140                           \\ 
                        & 4934.08                        & -0.160                          \\ 
                       & 5853.68                        & -0.908                          \\ 
                       & 6141.72                        & -0.032                          \\ 
                               & 6496.90                        & -0.407                          \\ 
                                \noalign{\smallskip}
                               \hline
                               \noalign{\smallskip}
\textbf{Magnesium \uproman{1}}          & 4353.13                        & -0.545                          \\ 
                        & 4572.38                        & -5.201                          \\ 
                       & 4702.99                        & -0.377                          \\ 
                        & 4731.35                        & -2.347                          \\ 
                               & 5528.41                        & -0.498                          \\ 
                               & 5711.09                        & -1.724                          \\ 
                                \noalign{\smallskip}
                               \hline
                               \noalign{\smallskip}
\textbf{Manganese \uproman{1}} & 4754.03                        & -0.080                          \\ 
                       & 4761.52                        & -0.274                          \\ 
                      & 4762.37                        & 0.304                           \\ 
                       & 4765.86                        & -0.086                          \\ 
                               & 4766.42                        & 0.105                           \\ 
                               & 4783.42                        & 0.044                           \\ 
                               & 6016.64                        & -0.181                          \\ 
                               & 6021.79                        & -0.054                          \\ 
\noalign{\smallskip}
\hline
\end{tabular}
\end{center}
\end{table}

%
%
%
%
% Figure: HFS/ISO
%
\begin{figure*}
   \includegraphics[width=0.9\textwidth ,center]{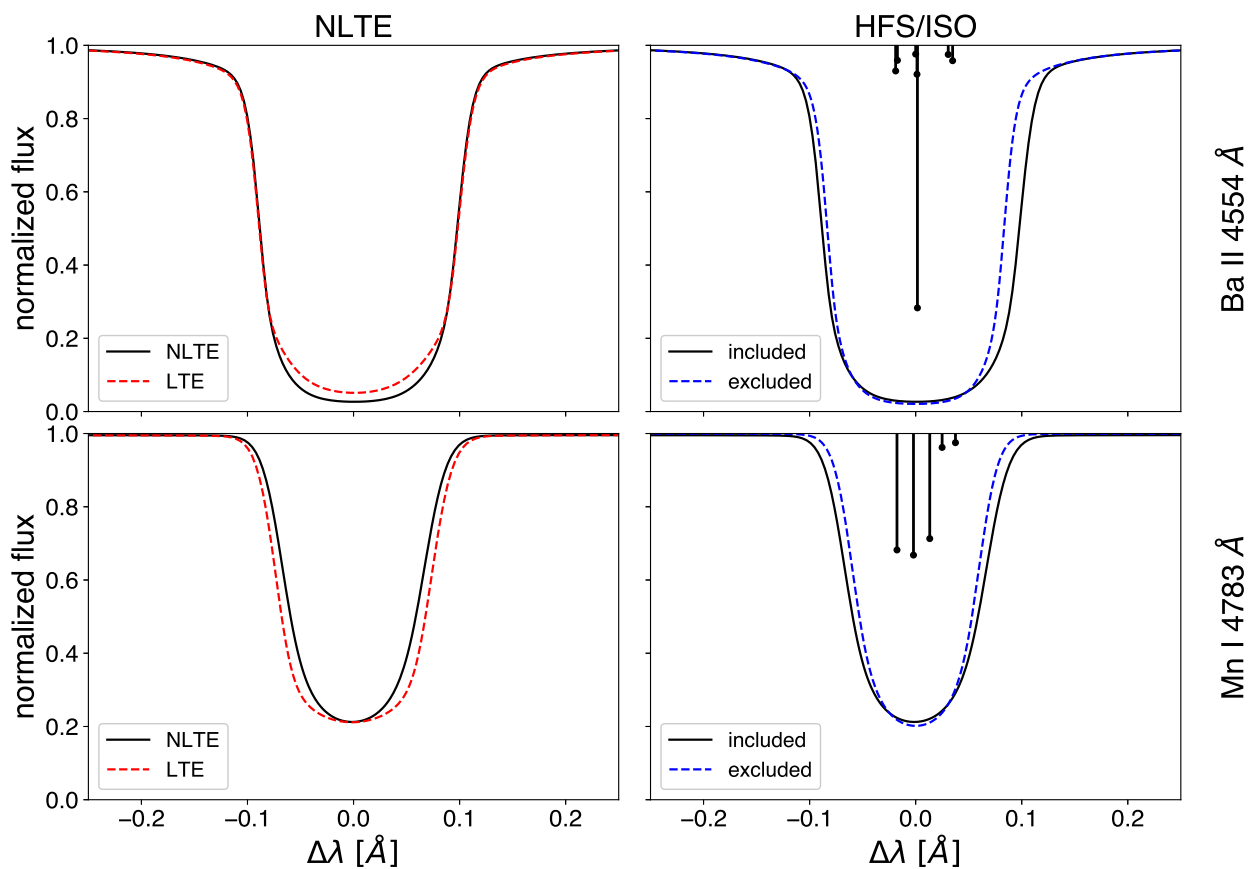}
   \caption[center]{Left panels: NLTE and LTE line profiles (left) for the Ba II 4554 $\AA$ (top) and Mn I 4783 $\AA$ lines (bottom) in the model with $T_{\rm eff} = 4382$ K, $\log g = 0.85$ dex, [Fe/H] $= -2$ dex, and $V_{\rm mic} = 2$ km/s. Right panels: The NLTE line profiles computed with and without hyperfine splitting (Mn) and isotopic shifts (Ba). The black vertical lines represent the individual components. Their length equals the weight of the respective component in the average of the full  profile.}
\label{f_NLTE-HFS}
\end{figure*}
\subsection{Atoms} \label{ss_atom}
We focus on three species: Ba \uproman{2}, Mg \uproman{1}, and Mn  \uproman{1}. The Mg atom has been presented and tested in \cite{Bergemann2017}. The Ba atom will be described in detail in Gallagher et al. (in prep). The model of Mn is the subject of Bergemann et al. (in prep.). The new Mn model is built upon the model atom presented in \cite{Bergemann2007}. The key difference with respect to earlier studies of these species is the inclusion of new quantum-mechanical rates of transitions caused by inelastic collisions with H atom \citep{Belyaev2017a,Belyaev2017,Rodionov2017}, as well as new quantum-mechanical photo-ionisation cross-sections for Mn I.

The model atoms contain the full representation of the energy levels, bound bound and bound-free radiative transitions in the system, as well as a description of the transitions, which are caused by collisions with H atom and free electrons. The Ba atom contains 111 levels and 394 radiatively-allowed transitions.  The Mg atom is composed of 86 states connected by 271 radiative transitions. Mn is represented by 281 levels with 1887 radiative transitions. The lines are represented by Voigt profiles with damping coefficients taken from \cite{Barklem2000}. The NLTE analysis is limited to those lines, which are typically observable in the optical spectra of individual stars and stellar populations. These lines and their atomic data are listed in Table \ref{t_lines}. The NLTE lines have a finer frequency quadrature to ensure the complete representation of their profiles.\\

For Ba and Mn atoms, we also include hyperfine splitting (HFS) and isotopic shifts (for Ba). The relative abundance of the five Ba isotopes (134,135,136,137,138) is taken to be solar. These effects have a very strong influence on the shape of line profiles even in very metal-poor model atmospheres. Fig.\ref{f_NLTE-HFS} compares the profiles of the Ba 4554 $\AA$  and Mn $5420$ lines in the model with $T_{\rm eff} = 4382$ K, $\log g = 0.85$ dex, [Fe/H] $= -2$ dex, and $V_{\rm mic} = 2$ km/s. The plots suggest that the accurate representation of the atomic properties of spectral lines is essential, in addition to NLTE effects, to obtain realistic spectra and correct abundances.\\
\begin{table*}[!htb]
\begin{center}
\caption{Stellar bins of the mock GC. L is the luminosity (given in solar units), M the mass of a star, T$_{\rm eff}$ is the effective temperature, g the surface gravity, R the radius. The weight represents the total number of stars in this bin. $V_{\rm mic}$ is the micro-turbulent velocity in km$/$s. The last column indicates the evolutionary stage of each bin: the main-sequence (MS),  turn-off (TO), sub-giant (SGB), red giant branch (RGB), and the horizontal branch (HB). All models have the metallicity of [Fe/H]$=-2$ dex.}
\label{t_model_atmos}
\setlength{\tabcolsep}{0.02\linewidth}
\begin{tabular}{r r c c c l c c r}
\hline
\noalign{\smallskip}
ID & L [$L_{\odot}$] & M [$M_{\odot}$] & T$_{
\rm eff}$ [K] & $\log_{10}(g)$ $[dex]$ & R [$R_{\odot}$] & Weight & $V_{\rm mic}$ $[km/s]$ & Stage  \\ 
\noalign{\smallskip}
\hline
\noalign{\smallskip}
1                              & 0.01                           & 0.26                           & 3928                           & 5.03                           & 0.26                           & 7.408E+05                      & 0.50                           & MS                              \\ 
2                              & 0.11                           & 0.53                           & 4728                           & 4.76                           & 0.50                           & 6.109E+04                      & 0.50                           & MS                              \\ 
3                              & 0.29                           & 0.62                           & 5460                           & 4.66                           & 0.61                           & 1.906E+04                      & 0.50                           & MS                              \\ 
4                              & 0.52                           & 0.67                           & 5853                           & 4.58                           & 0.70                           & 1.057E+04                      & 0.50                           & MS                              \\ 
5                              & 0.79                           & 0.70                           & 6126                           & 4.49                           & 0.79                           & 7.031E+03                      & 1.00                           & MS                              \\ 
6                              & 1.14                           & 0.73                           & 6333                           & 4.40                           & 0.89                           & 4.341E+03                      & 1.00                           & MS                              \\ 
7                              & 1.59                           & 0.75                           & 6474                           & 4.31                           & 1.00                           & 3.321E+03                      & 1.00                           & MS                              \\ 
8                              & 2.15                           & 0.76                           & 6545                           & 4.21                           & 1.14                           & 2.355E+03                      & 1.00                           & MS                              \\ 
9                              & 2.99                           & 0.77                           & 6480                           & 4.05                           & 1.37                           & 1.755E+03                      & 1.00                           & TO/SGB                          \\ 
10                             & 4.65                           & 0.78                           & 5800                           & 3.67                           & 2.14                           & 1.153E+03                      & 1.11                           & SGB                             \\ 
11                             & 10.3                          & 0.79                           & 5402                           & 3.20                           & 3.68                           & 5.584E+02                      & 1.26                           & RGB                             \\ 
12                             & 23.5                          & 0.79                           & 5269                           & 2.80                           & 5.83                           & 2.552E+02                      & 1.40                           & RGB                             \\ 
13                             & 48.2                          & 0.79                           & 5141                           & 2.45                           & 8.76                           & 1.198E+02                      & 1.52                           & RGB                             \\ 
14                             & 88.2                          & 0.79                           & 5013                           & 2.15                           & 12.5                          & 6.980E+01                      & 1.62                           & RGB                             \\ 
15                             & 139                         & 0.79                           & 4917                           & 1.91                           & 16.3                         & 4.771E+01                      & 1.69                           & RGB                             \\ 
16                             & 225                         & 0.79                           & 4798                           & 1.66                           & 21.8                          & 2.655E+01                      & 1.78                           & RGB                             \\ 
17                             & 370                         & 0.79                           & 4674                           & 1.40                           & 29.4                          & 1.903E+01                      & 1.87                           & RGB                             \\ 
18                             & 623                         & 0.79                           & 4529                           & 1.12                           & 40.6                          & 1.362E+01                      & 1.96                           & RGB                             \\ 
19                             & 1016                        & 0.79                           & 4382                           & 0.85                           & 55.4                          & 8.028E+00                      & 2.00                           & RGB                             \\ 
20                             & 1554                        & 0.79                           & 4251                           & 0.61                           & 72.8                          & 5.410E+00                      & 2.00                           & RGB                             \\ 
21                             & 47.5                          & 0.60                           & 9122                           & 3.33                           & 2.76                           & 2.827E+01                      & 1.80                           & HB                              \\ 
22                             & 48.3                          & 0.60                           & 7563                           & 3.00                           & 4.05                           & 2.827E+01                      & 1.80                           & HB                              \\  
\noalign{\smallskip}
\hline
\noalign{\smallskip}
\end{tabular}
\end{center}
\end{table*}

%---------------------------------------------------------
%
\subsection{Stellar atmosphere models} \label{ss_atmos}
A GC is a population that comprises of large number of FGKM-type stars. Known GCs in the MW have typical masses of 10$^4$ to 10$^6$ M$_{\rm Sun}$ \citep{Kruijssen2018} equivalent to 10$^5 .. 10^7$ stars. 
In principle, the integrated light of stellar clusters is subject to stochastic effects arising from the sampling of a finite number of stars from a given underlying mass function \citep[e.g.][]{Girardi1995,Bruzual2001,Fouesneau2010,Popescu2010}. However, for GCs with masses greater than about $10^5 M_\odot$, the impact of stochastic mass function (MF) sampling on abundance measurements from integrated light is small, typically less than 0.05 dex \citep{Larsen2017}. Here we model the integrated spectra as SSPs, assuming a smoothly populated MF, in which case the total mass is simply a scaling factor that does not affect spectral features. We model the SSP using a theoretical isochrone from the Dartmouth group \citep{Dotter2007} with an age of 11 Gyr, a metallicity of [Fe/H]=$-2.0$, and [$\alpha$/Fe]=$+0.4$. The Dartmouth isochrones list the relevant basic stellar parameters (mass $M$, effective temperature $T_\mathrm{eff}$, surface gravity $\log g$, and  bolometric luminosity $L/L_\odot$), as well as absolute magnitudes in several broad-band filters. These quantities are tabulated at 266 points along the isochrone, starting at the lower main sequence ($M=0.10 M_\odot$) and ending at the tip of the RGB. 
The stellar radii then follow from
$\log (R/R_\odot) = \frac{1}{2} \left[- \log g + \log g_\odot + \log(M/M_\odot)\right]$. 
However, to keep the calculations manageable we defined 20 bins along the isochrone that each contribute equally to the $V$-band luminosity. To define these bins, we assumed that the distribution of stellar masses follows the \cite{Salpeter1955} form, $dN/dM \propto M^{-2.35}$, and then calculated the mean stellar luminosity in the $V$-band ($\langle L_V \rangle_i$), weighted by the
assumed mass distribution, within the $i$-th bin. The physical parameters were then determined for each $\langle L_V \rangle_i$ by interpolation in the
full isochrone table, $T_{\mathrm{eff},i} = T_\mathrm{eff}(\langle L_V \rangle_i)$, $\log g_i = \log g(\langle L_V \rangle_i)$, etc. The Dartmouth isochrones only include stellar evolutionary phases up to the tip of the RGB, but we additionally include two points along the horizontal branch based on Hubble Space Telescope photometry of the Galactic GC NGC~6779, as described in \cite{Larsen2017}. The  stellar parameters for the resulting 22 bins are listed in Table \ref{t_model_atmos}, with masses in the range from $0.26$ to $0.79$ M$_{\rm Sun}$, $T_{\rm eff}$ in the range from $3900$ K (lower main-sequence, hereafter MS) to $\sim 10\,000$ K (horizontal branch) and $\log g$ in the range from $5$ (low MS) to $0.6$ (tip RGB) dex. The weight column indicates the number of stars per bin, and $V_{\rm mic}$  is the micro-turbulent velocity assumed for the modelling of the atmospheres and radiative transfer  \citep{Larsen2017}.\\

\begin{figure*}[!htb]
\centering
\includegraphics[width=0.99\textwidth ]{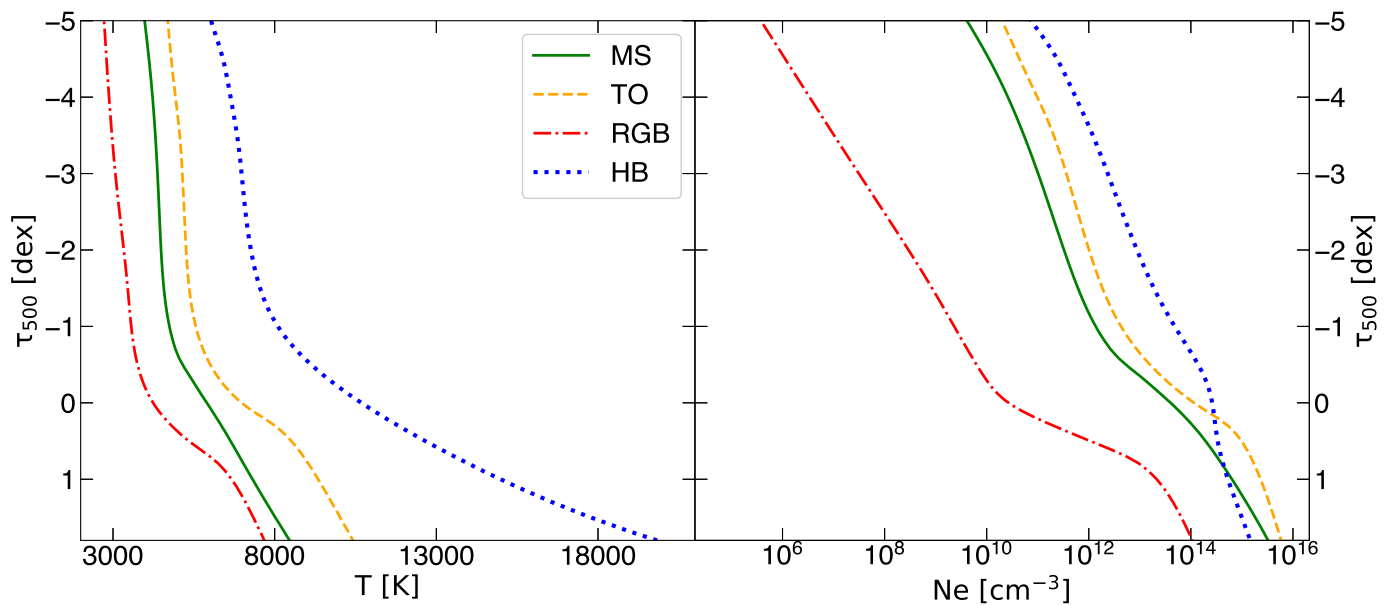}
\caption{The temperature and electron density profiles in selected model atmospheres from Table \ref{t_model_atmos}.}
\label{f_atmos}
\end{figure*}

For each set of stellar parameters in Table \ref{t_model_atmos}, we compute atmosphere models using the ATLAS12 code \citep{Sbordone2004,Kurucz2005}. ATLAS12 computes LTE, plane-parallel models for a given composition and microturbulent velocity, using opacity sampling. The physical structure of four model atmospheres for the key evolutionary stages is shown in Fig.\ref{f_atmos}. The atmospheres are $\alpha$-enhanced, [Mg/Fe]=$+0.4$ dex. The abundance ratios of Mn and Ba relative to Fe are assumed to be scaled solar. This is equivalent to a [(Ba, Mn)$/$H] $= -2$, [Mg$/$H] $=-1.6$ dex. The composition of the Dartmouth stellar models is given relative to the solar abundances of \cite{Grevesse1998}, which is the same abundance scale used for the ATLAS models.

Our reference solar abundances are $\log A(\rm{Ba}) = 2.11$ dex \citep{Holweger1974}, $\log A(\rm{Mn}) = 5.47$ dex \citep{Bergemann2007} and $\log A(\rm{Mg}) = 7.60$ dex \citep{Asplund2009}. We note that, in practice, the choice of the reference solar abundance is not important in the context of our work, as all calculations are performed for a very large dynamical range of elemental abundances ($\pm 2$ dex with respect to the central value, defined above) and we carefully explore the influence of the position of the central value on the NLTE abundance correction (see next section)
\subsection{Abundance corrections} \label{ss_abund_corr}
The equivalent width $W$ is a very useful parameter that characterises the total absorption in the spectral line.

The EW is defined as:
\begin{align}
W_i     &=    \int_{\lambda_1}^{\lambda_2} \left(1 -\frac{L_{i}}{L_{c,i}} \right)d\lambda \ ,
\label{eqw_def}
\end{align}
where $L_i$ refers to the luminosity of star $i$ in the wavelength range $[\lambda_1,\lambda_2]$ and $L_{c,i}$ is the continuum luminosity as a function of wavelength $\lambda$. $W$ varies with abundance and micro-turbulent velocity, as well as the stellar physical parameters $\log g$ and $T_\mathrm{eff}$.

We compute the grids of LTE, $W_L$, and NLTE, $W_N$, EWs for a range of different abundance values and for every spectral line from table \ref{t_lines}. The NLTE abundance grid covers a range of $\pm 2$ dex around the central value of [Mn,Ba/H] $= -2$ and [Mg/H] $= -1.6$, see Sect. \ref{ss_atom}, with a step size of 0.05 dex. This range encloses the complete Curve-of-Growth (CoG) for each line, including the linear part, the saturation part, and the damping part of the CoG and ensures that a closest matching EW can always be found within $W_N$ for a given reference $W_{L,r}$. Linear interpolation is applied between grid nodes of $W_N$ using the python library \texttt{scipy} \citep{Oliphant2007}. The result is a function for abundance, $A_N(W)$, which returns the NLTE abundance for a given EW. By subtracting the reference abundance $A_{L,r}$ from the one corresponding to the matching NLTE EW, the NLTE abundance correction,
\begin{align}
\label{eqw_abund_corr}
\Delta_{NLTE}  &=  A_N(W_{L,r}) -  A_{L,r} \ ,
\end{align} 
is obtained as a transition dependent quantity.\\
Since the shape and the depth of the resulting NLTE line profiles will not vary with abundance in the same way as they do in LTE, the corrections will depend on the chosen reference value. This is why we execute the calculations for a range of different reference abundance values in $W_L$, starting with the profile for the [(Ba,Mn)/Fe] $= 0$, [Mg/Fe] $= 0.4$ abundance ratios and varying them by $\pm 1$ dex with the same step size of 0.05 dex. Because our reference calculation is executed in LTE, we hereby simulate LTE abundance measurements and directly present the NLTE corrections for them. In other words, we present the NLTE abundance corrections for the abundance values, as an observer would measured them applying LTE models.\\

$\Delta_{\rm NLTE}$ is computed for every atom and for every atmospheric model. We can then determine the EW and the NLTE abundance corrections for every single star on its own or combine all line profiles to calculate the resulting integrated light EW for the whole GC.
In practice, we can choose between two methods of handling many stars at the same time.\\

\noindent Method 1: The continuum luminosity of each star can be expressed through the continuum luminosity of the first star, by multiplying a factor of $\xi$,
\begin{align} \label{eqw_xi}
L_{c,i}   &=  L_{c,1} \cdot \xi_i \ .
\end{align}
As presented in \cite{McWilliam2008}, and similar in \cite{Vazdekis1996}, the EW can be then calculated for the combined spectrum in the following manner, 
\begin{align} \label{eqw_cluster1}
W    &= \int_{\lambda_1}^{\lambda_2} \left(1 -\frac{\sum_i L_{i} }{\sum_i \xi_i L_{c,1}} \right)d\lambda \\
	   &=\frac{1}{\sum_i \xi_i} \cdot \sum_i \xi_i  \int_{\lambda_1}^{\lambda_2} \left( 1-\frac{L_{i}}{\xi_i L_{c,1}} \right) d\lambda\\
       &= \frac{1}{\sum_i \xi_i} \cdot \sum_i \xi_i W_i \ .
\end{align}
By calculating the single star equivalent widths, one obtains the EW of the whole GC by calculating the weighted average of the single star EWs with $\xi_i$ being the weights. Since we assumed earlier that the whole GC can be represented by 22 models, the sum in Eq.\ref{eqw_cluster1} contains 22 terms, in which their $\xi$ values need to by multiplied by the weight column of Table \ref{t_model_atmos}, which represents the total number of stars $N_i$ in each bin. By introducing the flux $F$ in the same notation as the luminosity, one obtains an equation for the EW of the total GC:
\begin{align} \label{eqw_cluster_final1}
W    &= \frac{\sum_i N_i \cdot R_i^2 F_{c,i} W_i}{\sum_i N_i \cdot R_i^2 F_{c,i}} \ ,
\end{align}
where $R_i$ is the radius of the stars in each bin.\\

\noindent Method 2: It is also possible to first combine the individual fluxes of each star and calculate one total EW out of the combined integrated light spectrum. 
\begin{align} \label{ew_lum2}
L(\lambda)    &=  \sum_i N_i \cdot 4\pi R_i^2 F_i(\lambda) \ .
\end{align}
Under the condition that the resulting line profile $L(\lambda)$ is broad enough, the continuum luminosity is contained in $L$ as being the outermost part of the line profile. This is also valid for the continuum flux in Eq.\ref{eqw_cluster_final1}. The EW is then calculated using Eq.\ref{eqw_def}.

By executing each of the two described methods, one obtains an integrated light EW for the entire GC, providing the possibility of determining integrated light NLTE abundance corrections according to Eq.\ref{eqw_abund_corr}. The question of which method to choose may vary from case to case. If the NLTE corrections for the individual stellar atmosphere models are computed anyway, re-using the EW data might be more efficient than adding up the spectra, especially for large grids of abundances and models. On the other hand side, method 2 produces the complete integrated light line profile as a side-product, which might also be of interest in the context of some research studies. Also, Eq.\ref{eqw_xi} holds when the star-to-star variations in the continuum shape across the feature of interest are negligible. This is a good approximation for individual lines, but might not be adequate for broader features, such as molecular bands, in which case one may have to rely on method 2. \\
In all cases, the EWs are calculated using the numerical integration, which is more precise the Simpsons method \citep{Press1986}. To obtain the EW the flux is first normalised to its value corresponding to the lowest wavelength available, which is considered to be the continuum flux. In this paper, the maximum wavelength offset from the line centre is 0.3 $\AA$, which is fully sufficient to describe the profiles of all selected lines. The normalised quantity is then inverted, by subtracting it from one, and integrated over wavelength.
%
%--------------------------------------------------------
%
\section{Results} \label{s_results}
\begin{figure*}
\includegraphics[width=\textwidth]{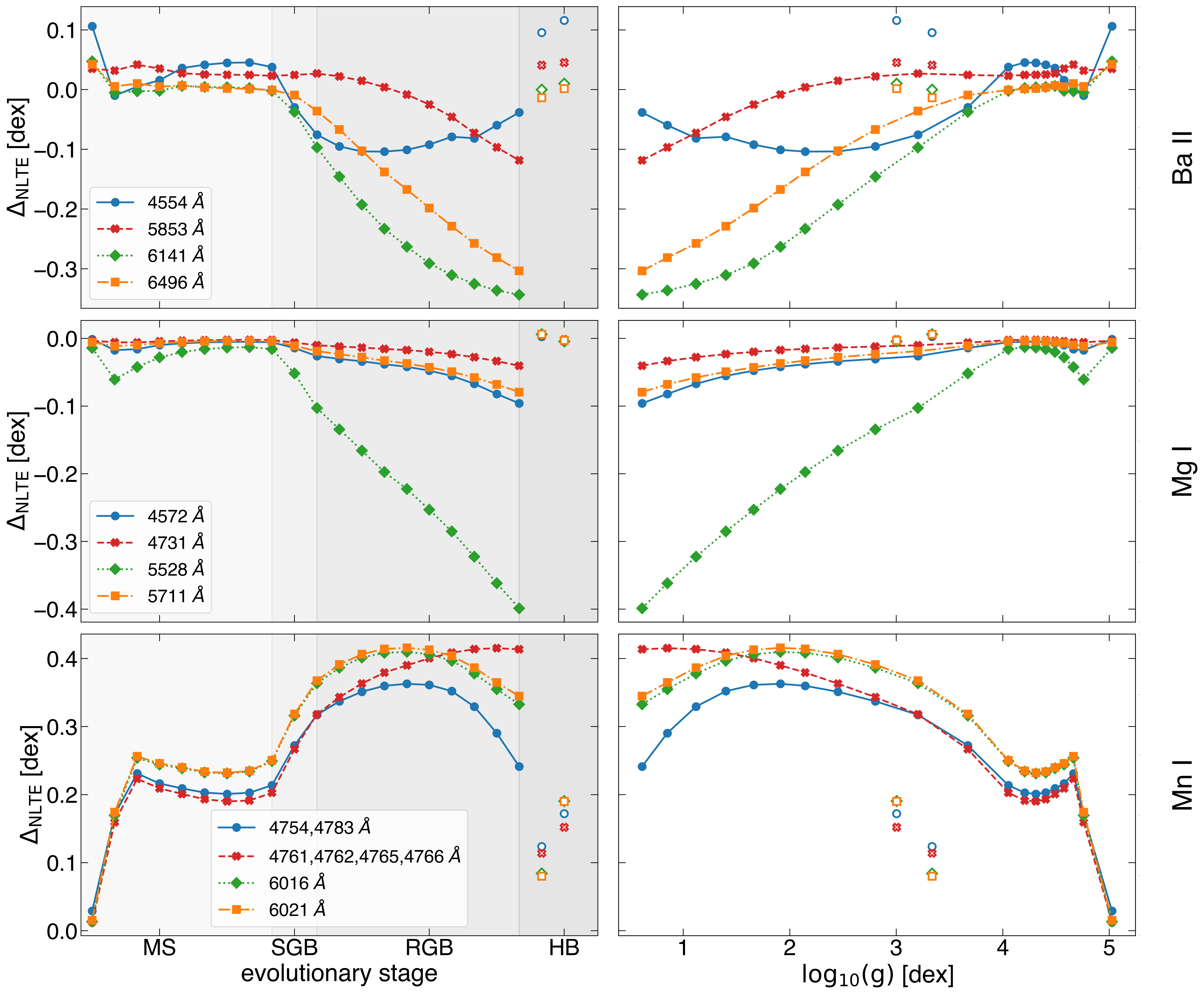}
\caption{Ba, Mg, and Mn NLTE abundance corrections as a function of the evolutionary stage and surface gravity for all models listed in Table \ref{t_model_atmos}. For Mn, the lines are grouped in multiplets and the average NLTE correction is shown. The LTE reference is chosen to be a common measured abundance in GCs, that is $\rm [Ba/Fe]=-0.2$, $\rm [Mn/Fe]=-0.4$, $\rm [Mg/Fe]=+0.4$. The empty markers represent the horizontal branch stars.}
\label{f_corr_evol}

\end{figure*}
In what follows, we present the results of our NLTE calculations for single star and integrated light models. Section \ref{ss_results_indiv_star} presents the NLTE abundance corrections for all transitions listed in Table \ref{t_lines} for individual stars that make up the SSP. Section \ref{ss_results_iteg_light} presents the corrections for the integrated light model of the SSP.
\subsection{Individual stars} \label{ss_results_indiv_star}
The NLTE abundance corrections for Ba, Mg, and Mn for all model atmospheres in the SSP are shown in Fig.\ref{f_corr_evol}, where the results are plotted against surface gravity. On this figure we present the corrections derived for an LTE reference abundance, which is typically measured in globular clusters with similar metallicity, namely: [Mg/Fe] $= 0.4$ \citep{Sneden1997a}, [Mn/Fe] $= -0.4$ \citep{Sobeck2006a}, [Ba/Fe] $= -0.2$ \citep{Larsen2018}\footnote{Although, we note that the scatter of [Ba/Fe] measurements at [Fe/H]$\sim -2$ dex is typically very large \citep[][Fig.9]{Larsen2018}.}. However, as described earlier, we compute NLTE corrections for a large grid of different LTE references. For the results for other reference abundances, we refer to Table \ref{t_intLight_complete} in Appendix \ref{appendix_lte_ref}.\\

The NLTE corrections for Ba \uproman{2} vary from $\approx 0$ dex in main sequence and horizontal branch stars to $\approx -0.3$ dex for the $6496$ \AA\ transition in red giants. However, there is also a significant variation among the individual Ba \uproman{2} lines. The weakest line at 5853 $\AA$ shows only negligible NLTE effects for the MS and sub-giants, but the (negative) NLTE correction increases linearly with surface gravity up to the tip of the RGB. In the HB models, the NLTE correction is very small. The other three diagnostic Ba \uproman{2} lines show a non-negligible NLTE effect already in some MS models, and the NLTE corrections are also very sensitive to the input LTE abundance see Sec.\ref{ss_ref_depend} and Appendix \ref{appendix_lte_ref}. The most striking variation is seen towards the middle - upper part of the RGB, where the NLTE correction for the 4554 $\AA$ line is only $-0.1$ dex, but amounts to $-0.35$ dex for the $6141$ and $6496 \ \AA$ lines. Ba is a collision-dominated ion, hence its statistical equilibrium is predominantly controlled by strong line scattering, that is, radiative pumping in the deeper atmospheric layers and photon losses higher up in the atmosphere \citep{Mashonkina1996, Mashonkina1999, Short2006}. This can be understood using the concept of the departure coefficient $b$ of the energy levels involved in a transition, where $b = n_{\rm i \ NLTE}/n_{\rm i \ LTE}$, the ratio of the atomic number density of an energy level $i$ in NLTE relative to its LTE value. The line source function $S_{\nu}$ is related to the Planck function $B_{\nu}$, as $S_{\nu}/B_{\nu} \sim b_j/b_i$ \citep{Bergemann2014a}. For the strong Ba lines, $b_i > b_j$ in the line core forming region, resulting in a sub-thermal source function ($S_{\nu}<B_{\nu}$), and, hence, negative NLTE abundance corrections.

Comparing our estimates with the earlier studies of NLTE effects in Ba \citep{Korotin2015}, we find that the agreement is very good. The corrections for the RGB models, $L > 10~L_{\odot}$, are larger (that is, more negative) for subordinate lines at 5853, 6141, and 6496 $\AA$, while they are more moderate for MS stars. For example, for a typical MS model our average correction for the Ba \uproman{2} 6496 $\AA$ line is $\sim 0$ dex. Comparing this result with Korotin's Fig.A1, they also obtain a correction of $\sim 0$ dex for a model with $T_{\rm eff} \approx 5000 $K, $\log g = 4$ dex, $V_{\rm mic} = 2$ km/s, the metallicity of $-2$ dex, and a reference [Ba/Fe] $= -0.3$ dex. For the RGB stars, the results are also very similar. Korotin et al. find the NLTE corrections of $\sim -0.25$ dex for the model with $T_{\rm eff} \approx 4500$ K and $\log g = 1$ dex, which is in a very good agreement with our results of $-0.2$ to $-0.3$ for similar RGB models (Fig.\ref{f_corr_evol}). The residual difference is likely caused by the difference in the model atoms: Korotin et al. assumed Drawin's hydrogenic-like cross-sections for the Ba$+$H collisions, whereas we employ the new ab-initio quantum-mechanical data from \cite{Belyaev2017a,Belyaev2017}. Also the reference abundances and the micro-turbulent velocities of their models are slightly different.\\

The Mg lines typically show negative NLTE effects, in line with the earlier studies \citep{Merle2011,Osorio2015,Bergemann2017}. The NLTE abundance corrections for the bluer lines at 4572 $\AA$ and 4731 $\AA$ are very modest, and do not exceed $-0.1$ dex. On the other hand, the strong line at $5528 \ \AA$ is very sensitive to NLTE and show the NLTE abundance correction of $-0.4$ dex in the models of stars on the upper RGB, $\log g < 1.5$. NLTE effects in Mg are driven by a competition of several processes, but, in particular, the very higher-excitation lines at 5711 and 5528 \AA~ (with the lower energy level at $\sim 4.35$ eV) experience NLTE line darkening, caused by the line scattering and downward electron cascades from above \citep{Bergemann2017}, generally causing $b_i$ to exceed $b_j$ that leads to sub-thermal $S_{\nu}$.

Our results are consistent with the recent theoretical calculations by \cite{osorio2016} (see their Fig.2), who also obtain large and negative NLTE corrections for this subordinate line in metal-poor models. For example, contrasting our RGB model \# 14 ($\log g = 2.15$ dex) with the table 2 in Osorio \& Barklem, we find an agreement to within $0.05$ dex for all lines that overlap with their sample. We also confirm that the NLTE corrections for the 5711 line are very small, close to zero, for stars of all evolutionary phases.\\

The only study that explored the NLTE effects in Mn is that of \citet{Bergemann2007,Bergemann2008}. However, in that study we did not consider NLTE effects in RGB stars, neither did we have accurate atomic data for Mn$+$H collisions or photo-ionisation. Qualitatively, we confirm our earlier results that NLTE corrections for Mn are typically non-negligible and are mostly positive. This is the consequence of the atomic structure of Mn I: with its large photo-ionisation cross-sections and multiple energy levels with ionisation thresholds in the near-UV and optical, Mn I is a typical over-ionisation specie. The  positive NLTE effects are driven by strong radiative over-ionisation that causes significant changes in the line opacity compared to LTE \citep{Bergemann2008}.

The detailed behaviour of the NLTE corrections as a function of $\log g$ or $T_{\rm eff}$ is slightly different. In this work, we find the maximum NLTE-LTE differences for the lines of the $\rm a \ ^4D - z \ ^4F^o$ multiplet at 4761 - 4766 \AA, whereas the lines of multiplet $\rm z \ ^8P^o -e \ ^8S$ (4754, 4783, 4823 $\AA$), as well as the high-excitation lines of multiplet $\rm z \ ^6P^o - e \ ^6S$ (6013, 6021 $\AA$), show corrections of the order 0.2 dex for main-sequence to maximum $0.4$ dex for the RGB stars.

\subsection{The sensitivity to the absolute abundance}
\label{ss_ref_depend}
\begin{figure}
   \includegraphics[width=0.5\textwidth,center]{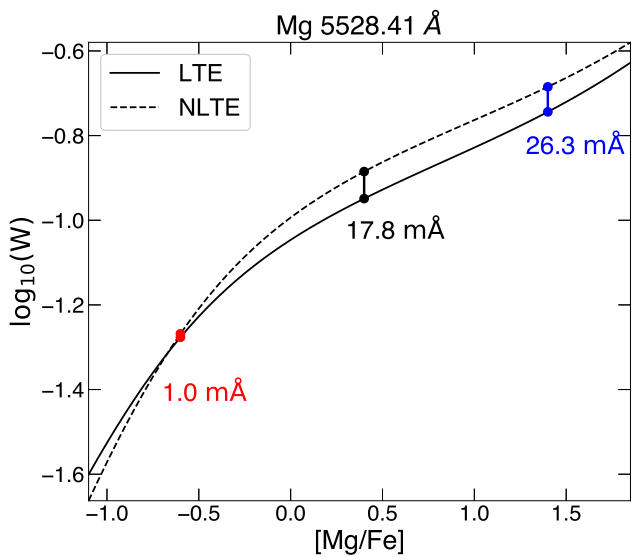}
    \caption{NLTE/LTE CoG comparison. The vertical axis is scaled in logarithmic units of \AA. Marked are the central reference abundance (black), as well as the 1 dex increased (blue) and decreased (red) ones. The difference between the NLTE and LTE equivalent width for each reference point is written inside the plots.}
    \label{f_cog_ref_comp}
\end{figure}
In this section we present the NLTE corrections for different reference abundances. Additional information can be found in Appendix \ref{appendix_lte_ref}. 

For some transitions, the results vary strongly with the reference (LTE) abundance, even within the same model atmosphere. The most sensitive lines are the Ba \uproman{2} 6496 $\AA$, Mg \uproman{1} 5528 $\AA$, and Mn \uproman{1} 4754 $\AA$ lines. Their NLTE corrections for the RGB models change dramatically, depending on the choice of the zero point value. On the other hand, the NLTE corrections are typically small for the main-sequence models. For Mn and Ba, lowering the reference abundances leads to less negative NLTE abundance corrections, while increasing the reference abundance yields more negative NLTE corrections.\\

We can understand the origin of this peculiar behaviour by exploring the NLTE and LTE CoGs. Since the most striking deviations are seen for the RGB stars, Fig.\ref{f_cog_ref_comp} shows the EW as a function of the abundance for the Mg \uproman{1} 5528 $\AA$ line in the model atmospheres \# 18 ($\rm{T_{eff}} = 4382$,  $\rm{log g}=0.85$). 
In the regime of low Mg abundance, [Mg/Fe] $\lesssim -0.5$, the NLTE and LTE CoG curves are very similar and thus the NLTE corrections will be close to zero. For the lowest reference abundance of [Mg/Fe]$ = -0.6$, the difference between the NLTE and LTE EWs is only 1 $m\AA$, compared to 17.8 $m\AA$ computed using [Mg/Fe] $= 0.4$ dex. However, with increasing the Mg abundance, the curves diverge more and more, which reflects the growing importance of NLTE line scattering with increasing line strength. Consequently, $S_{\nu}$ becomes sub-thermal. At [Mg/Fe] $\approx 0$ dex, the CoG enters the saturated regime, where a small change in the EW corresponds to a large change in the abundance. 

For Ba lines, this behaviour is qualitatively very similar and has the same origin. However, for Mn, the NLTE CoG is located below the LTE curve leading to positive NLTE corrections.
\subsection{Integrated light} \label{ss_results_iteg_light}
%
%
% Table: integrated Light 
%
\begin{table}
\begin{center}
\caption{Integrated Light NLTE abundance corrections according to method 1 in section \ref{ss_abund_corr}. References are mentioned in the first column.}
\label{t_intLight}
\setlength{\tabcolsep}{0.06\linewidth}
\begin{tabular}{l c c}
\hline
\noalign{\smallskip}
Element/Ion & Wavelength [$\AA$] & $\rm{\Delta_{NLTE}}$  \\ 
\noalign{\smallskip}
\hline
\noalign{\smallskip}
\textbf{Barium \uproman{2}}             & 4554.03                        & -0.023                          \\ 
$\rm{[Ba/Fe] = -0.2}$          & 4934.08                        & -0.098                          \\ 
                               & 5853.68                        & -0.032                          \\ 
                               & 6141.72                        & -0.193                          \\ 
                               & 6496.90                        & -0.156                          \\ 
                               \noalign{\smallskip}
                               \hline
                               \noalign{\smallskip}
\textbf{Magnesium \uproman{1}}          & 4353.13                        & -0.051                          \\ 
$\rm{[Mg/Fe] = +0.4}$          & 4572.38                        & -0.028                          \\ 
                               & 4702.99                        & -0.084                          \\ 
                               & 4731.35                        & -0.019                          \\ 
                               & 5528.41                        & -0.152                          \\ 
                               & 5711.09                        & -0.036                          \\ 
                               \noalign{\smallskip}
                               \hline
                               \noalign{\smallskip}
\textbf{Manganese \uproman{1}} & 4754.03                        & 0.287                           \\ 
$\rm{[Mn/Fe] = -0.4}$          & 4761.52                        & 0.328                           \\ 
                               & 4762.37                        & 0.341                           \\ 
                               & 4765.86                        & 0.333                           \\ 
                               & 4766.42                        & 0.337                           \\ 
                               & 4783.42                        & 0.292                           \\ 
                               & 6016.64                        & 0.329                           \\ 
                               & 6021.79                        & 0.337                           \\ 
\noalign{\smallskip}
\hline
\end{tabular}
\end{center}
\end{table}

The integrated light NLTE abundance corrections for Ba, Mg, and Mn are shown in Table \ref{t_intLight}. For consistency, we apply both methods described in Sect.\ref{s_methods}. However, comparing the results, we find no significant deviations between the results obtained with both methods, hence we only present the values computed using method 1.

Similar to the individual stars, also the integrated light NLTE corrections depend on the chosen reference (LTE) abundance. We refer the reader to Appendix, where Table \ref{t_intLight_complete} shows the NLTE corrections for a range of different abundances. In this section, we discuss the results for the typical abundance ratios.

The tendency for LTE to over-estimate (Mg,Ba), respectively, under-estimate (Mn) abundances clearly stands also in the integrated light results (Table \ref{t_intLight}). The NLTE corrections for Mg \uproman{1} lines are very small, and tend to be negative. In particular, the 5528 \AA\ line appears to be most sensitive to the NLTE effects. The situation is qualitatively similar for Ba \uproman{2}: we find the smallest NLTE abundance corrections of the order $\sim -0.03$ dex for the 4554 \AA\ and 5853 \AA\ Ba lines, whereas the high-excitation stronger lines at 6141 and 6496 \AA\ show the NLTE corrections of $\approx -0.3$ dex. The NLTE corrections for all Mn \uproman{1} lines are very similar and exceed $0.3$ dex, suggesting that the LTE approach would significantly under-estimate Mn abundances in stars and stellar populations.

It should be stressed that even though the NLTE abundance corrections depend on the atomic properties of the line, they do much less so in the integrated light compared to individual stars. For example, the NLTE corrections for the Mg \uproman{1} 5528 $\AA$ line change from $0$ on the main-sequence to  $-0.35$ dex in tip RGB models (Fig.\ref{f_corr_evol}). On the other hand, the integrated light correction for the line amounts to $-0.15$ dex only.  

In order to qualitatively estimate the sensitivity of our results to the treatment of the horizontal branch, we additionally calculated the integrated light NLTE corrections without taking models \# 21 and 22 into account. The resulting difference to the original values then gauges their influence on the NLTE corrections of the GC. However, the differences are in the order of only $0.001$ dex and never exceed $0.01$ dex for all considered lines. Hence, in our case the HB stars have no significant impact on the abundance corrections. Furthermore, we excluded the model \# 1 to qualitatively investigate the influence of a 'top-heavy' IMF on the abundance corrections. Here we found that the differences are on average $0.01$ dex for the Mn I  lines, and even less for the other species. They do not exceed $0.02$ dex.
\section{Discussion}
It is interesting to consider these results in the context of previously published integrated-light studies of globular clusters. Similarly to studies of individual metal-poor halo stars, integrated-light observations of metal-poor GCs typically yield significantly depleted Mn abundances of $\mathrm{[Mn/Fe]}\simeq-0.4$ (e.g. \citealt{Colucci2012,Larsen2018}). Our results suggest that this can be explained mostly as a NLTE effect, with the actual $\mathrm{[Mn/Fe]}$ ratios most likely being close to solar. our models predict that there is no significant variation in the NLTE abundance correction between the individual Mn \uproman{1} lines, although the redder lines at 6016,6021 \AA\ have somewhat, $\sim 0.05$ dex, larger NLTE correction compared to the bluer line at 4754 \AA. This is consistent with the LTE abundance measurements in the integrated light spectra \citep[e.g.][Table A.7]{Larsen2018}, which do not detect strong differential LTE abundance variations among the optical Mn lines. \\

Another element of interest is Mg, which tends to be enhanced in metal-poor field stars at a similar level ($\mathrm{[Mg/Fe]}\approx+0.4$) as other $\alpha$-elements (e.g. \citealt{Venn2004}). However, some integrated-light GC spectra show much less enhanced $\mathrm{[Mg/Fe]}$ ratios, which has been proposed to be related to the Mg-Al anti-correlation (e.g. \citealt{Colucci2009,Larsen2018}). Our results indicate that these Mg deficiencies are unlikely to be caused by NLTE effects. In fact, the NLTE corrections for Mg (although small for all Mg \uproman{1} lines, but the line at 5528 \AA) tend to be negative. We find a significant difference between the 5528 and 5711 \AA\ lines, such that the NLTE correction for the former is much larger and negative compared to the latter. However, this behaviour is not detected in LTE modelling \citep{Larsen2018}. In contrast, LTE abundance measurements suggest that abundance derived from the 5528 \AA\ line is too low compared to that computed from the 5711 line. At this stage, we do not have an explanation to this discrepancy. Our NLTE corrections are consistent with other 1D NLTE studies, but it is possible that the effects of 3D and convection play a non-negligible role at this metallicity. On the other hand, it is also possible that such a direct comparison to LTE abundances measurements is not meaningful, as those employ LTE metallicities that are also biased in the low-[Fe/H] regime \citep{Ruchti2013}. Also one has take into account that the NLTE corrections are sensitive to the choice of the LTE abundance, and according to our results (Fig. \ref{f_indv_corr_logg}), the Mg 5528 \AA\ line is particularly sensitive to this effect. \\

For Ba, the corrections in Table \ref{t_intLight} suggest that integrated-light [Ba/Fe] ratios may be overestimated by $\simeq0.1$ dex at metallicity [Fe/H]$=-2$. \citet{Larsen2018} found that the [Ba/Fe] ratios of metal-poor GCs in dwarf galaxies appeared to be somewhat higher ($\simeq0.2$ dex) than those seen in Milky Way halo stars \citep{Ishigaki2013}, but this comparison is made more difficult by the fact that no NLTE corrections were applied to the field stars. LTE abundance measurements of Ba in the integrated light spectra of GCs are not conclusive on whether there is a significant dispersion between the individual Ba \uproman{2} lines or not. If we consider the two metal-poor GSs from \citep{Larsen2018}, NGC 7078 ([Fe/H]$= -2.4$) and NGC 7099 ([Fe/H]$= -2.35$), we find that our NLTE corrections would not improve the consistency between different lines. However, similar to Mg, the comparison with the LTE abundance measurements is likely not meaningful at this stage, as there are many latent parameters in the method that are not consistent in our work and in previous LTE studies. 
\section{Conclusions} \label{s_discussion}
In this work, we introduce a new method to determine NLTE abundance corrections from synthetic integrated light spectra of simple stellar populations. We use two methods: the single-star measurements of equivalent widths and a linear superposition of their spectra. 

We present NLTE abundance corrections for Ba, Mg, and Mn for an old mono-metallic metal-poor globular cluster at [Fe/H] $=-2$. We find that the departures from LTE occur not only in the analysis of individual stellar lines, but also in the combined light spectra. They are mostly significant for Ba \uproman{2} and Mn \uproman{1} ions. For the optical lines of Mn I, the average NLTE abundance correction is $+0.3$ dex. Ba II and Mg I lines show significant differential effects: the low-excitation lines of both ions are almost unaffected by NLTE, whereas the high-excitation lines show NLTE abundance corrections of the order $\sim -0.15$ dex. Our results emphasise the need for taking NLTE effects into account in the analysis of chemical composition of stellar populations.
 
Following the studies of individual stars, it is especially interesting to investigate the metallicity dependence of the NLTE corrections. Also a comparison of GCs with different ages could be carried out in this context. Furthermore, the analysis of other atoms, for example Na or other $\alpha$-elements like O or Ca, would be straightforward and of a huge interest alongside with a larger set of spectral lines. Also a big step towards a most realistic and reliable abundance determination is the use of 3D convective simulations of stellar atmospheres, which will be considered in our future work.

\section*{Acknowledgements}
We are very thankful to our referee Dr. Ian Short for many constructive suggestions, which have helped to improve the paper. We acknowledge  support by the Collaborative Research centre SFB 881 (Heidelberg University) of the Deutsche Forschungsgemeinschaft (DFG, German Research Foundation). 
%
%
% Bib
%
\bibliographystyle{aa}
\bibliography{references}

\clearpage
%
% Appendix
%
\onecolumn
\appendix
\section{LTE reference dependence} \label{appendix_lte_ref}
\begin{figure*}[!hb]
   \includegraphics[width=\textwidth,center]{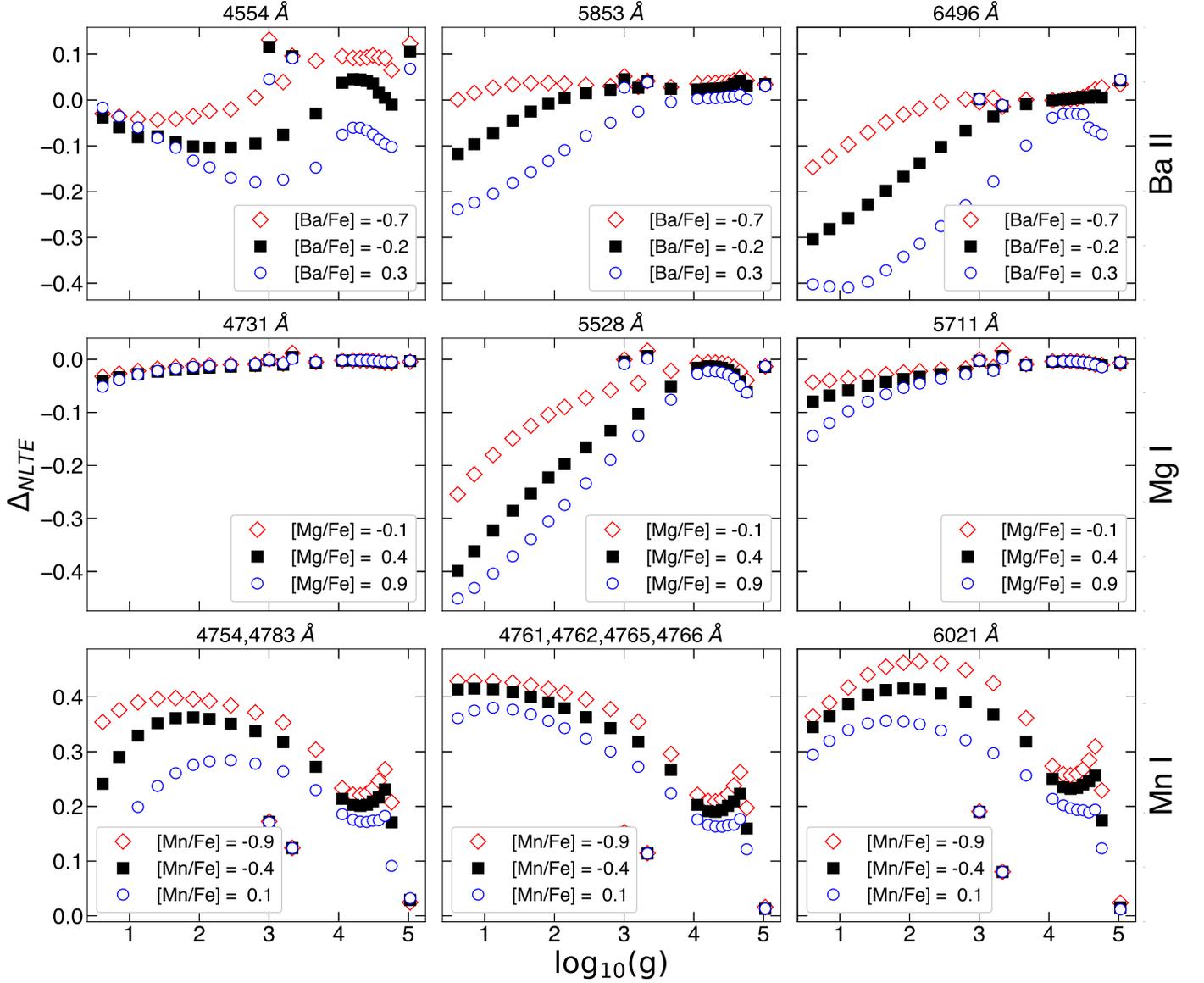}
    \caption{Ba, Mg, and Mn NLTE abundance corrections as a function of surface gravity for all models listed in Table \ref{t_model_atmos}. For Mn the lines are grouped in multiplets and the average NLTE correction is shown. The black '$\Box$' represent the corrections for a typically measured LTE abundance as reference. The blue '$\circ$' and the red '$\Diamond$' show the corrections for an 0.5 dex increased and decreased reference abundance, respectively.}
    \label{f_indv_corr_logg}
\end{figure*}

The NLTE integrated light abundance corrections are tabulated for a grid of different reference abundances in Table \ref{t_intLight_complete}. The reference dependence investigated in Sec.\ref{ss_ref_depend} is visualised in Fig.\ref{f_indv_corr_logg}.
\begin{table*}[!htb]
\begin{center}
\caption{Integrated Light NLTE abundance corrections for a grid of reference abundances. Each column shows the corrections for a different reference listed in the header. Here 'ref' means the central LTE reference abundances ($\rm{[Ba,Mn/Fe] = 0}$, $\rm{[Mg/Fe] = 0.4}$), the numbers represent the applied offset, i.e. [Atom/Fe] = reference + offset.}
\label{t_intLight_complete}
\setlength{\tabcolsep}{0.0125\linewidth}
\begin{tabular}{l r r r r r r r r r r r r}
\hline
\noalign{\smallskip}
Element/Ion & $\lambda$ [$\AA$] & -1.0 & -0.8 & -0.6 & -0.4 & -0.2 & ref & +0.2 & +0.4 & +0.6 & +0.8 & +1.0  \\ 
\noalign{\smallskip}
\hline
\noalign{\smallskip}
\textbf{Ba \uproman{2}}             & 4554.03                        & 0.067                          & 0.044                          & 0.024                          & 0.002                          & -0.023                         & -0.049                         & -0.075                         & -0.094                         & -0.104                         & -0.105                         & -0.097                          \\ 
                  & 4934.08                        & 0.015                          & -0.016                         & -0.045                         & -0.071                         & -0.098                         & -0.125                         & -0.150                         & -0.170                         & -0.177                         & -0.170                         & -0.150                          \\ 
                               & 5853.68                        & 0.043                          & 0.033                          & 0.017                          & -0.005                         & -0.032                         & -0.063                         & -0.095                         & -0.128                         & -0.159                         & -0.187                         & -0.214                          \\ 
                               & 6141.72                        & -0.072                         & -0.100                         & -0.130                         & -0.161                         & -0.193                         & -0.222                         & -0.250                         & -0.280                         & -0.308                         & -0.333                         & -0.346                          \\ 
                               & 6496.90                        & -0.019                         & -0.045                         & -0.078                         & -0.116                         & -0.156                         & -0.195                         & -0.233                         & -0.272                         & -0.309                         & -0.344                         & -0.373                          \\ 
\noalign{\smallskip}
\hline
\noalign{\smallskip}
 \textbf{Mg \uproman{1}}          & 4353.13                        & 0.004                          & -0.009                         & -0.021                         & -0.032                         & -0.042                         & -0.051                         & -0.057                         & -0.060                         & -0.059                         & -0.055                         & -0.049                          \\ 
               & 4572.38                        & -0.011                         & -0.020                         & -0.025                         & -0.028                         & -0.029                         & -0.028                         & -0.026                         & -0.025                         & -0.024                         & -0.023                         & -0.022                          \\ 
                               & 4702.99                        & 0.000                          & -0.018                         & -0.037                         & -0.054                         & -0.071                         & -0.084                         & -0.092                         & -0.094                         & -0.091                         & -0.083                         & -0.073                          \\ 
                               & 4731.35                        & 0.004                          & -0.006                         & -0.013                         & -0.017                         & -0.019                         & -0.019                         & -0.018                         & -0.017                         & -0.016                         & -0.016                         & -0.017                          \\ 
                               & 5528.41                        & -0.022                         & -0.048                         & -0.075                         & -0.102                         & -0.128                         & -0.152                         & -0.170                         & -0.180                         & -0.179                         & -0.169                         & -0.152                          \\ 
                               & 5711.09                        & -0.004                         & -0.015                         & -0.023                         & -0.029                         & -0.033                         & -0.036                         & -0.040                         & -0.045                         & -0.051                         & -0.058                         & -0.065                          \\ 
\noalign{\smallskip}
\hline
\noalign{\smallskip}

 \textbf{Mn \uproman{1}}                             & 4754.03                        & 0.338                          & 0.325                          & 0.309                          & 0.287                          & 0.261                          & 0.228                          & 0.191                          & 0.151                          & 0.109                          & 0.066                          & 0.021                           \\ 
                                & 4761.52                        & 0.361                          & 0.353                          & 0.341                          & 0.328                          & 0.312                          & 0.294                          & 0.272                          & 0.246                          & 0.217                          & 0.183                          & 0.146                           \\ 
                               & 4762.37                        & 0.371                          & 0.364                          & 0.354                          & 0.341                          & 0.324                          & 0.302                          & 0.274                          & 0.241                          & 0.203                          & 0.161                          & 0.116                           \\ 
                               & 4765.86                        & 0.363                          & 0.356                          & 0.346                          & 0.333                          & 0.317                          & 0.299                          & 0.275                          & 0.248                          & 0.215                          & 0.178                          & 0.138                           \\ 
                               & 4766.42                        & 0.366                          & 0.359                          & 0.350                          & 0.337                          & 0.321                          & 0.302                          & 0.276                          & 0.246                          & 0.211                          & 0.171                          & 0.128                           \\ 
                               & 4783.42                        & 0.348                          & 0.335                          & 0.316                          & 0.292                          & 0.262                          & 0.227                          & 0.187                          & 0.146                          & 0.102                          & 0.058                          & 0.013                           \\ 
                               & 6016.64                        & 0.369                          & 0.359                          & 0.346                          & 0.329                          & 0.311                          & 0.290                          & 0.265                          & 0.236                          & 0.202                          & 0.162                          & 0.117                           \\ 
                               & 6021.79                        & 0.373                          & 0.364                          & 0.352                          & 0.337                          & 0.320                          & 0.299                          & 0.273                          & 0.241                          & 0.202                          & 0.158                          & 0.108                           \\ 
\noalign{\smallskip}
\hline
\end{tabular}
\end{center}
\end{table*}

\clearpage
\section{HRD diagrams}  \label{appendix_evol_plots}
\begin{figure*}[!hb]
%
% Figure: HRD
%
\begin{subfigure}{.5\textwidth}
   \includegraphics[width=0.99\textwidth ,center]{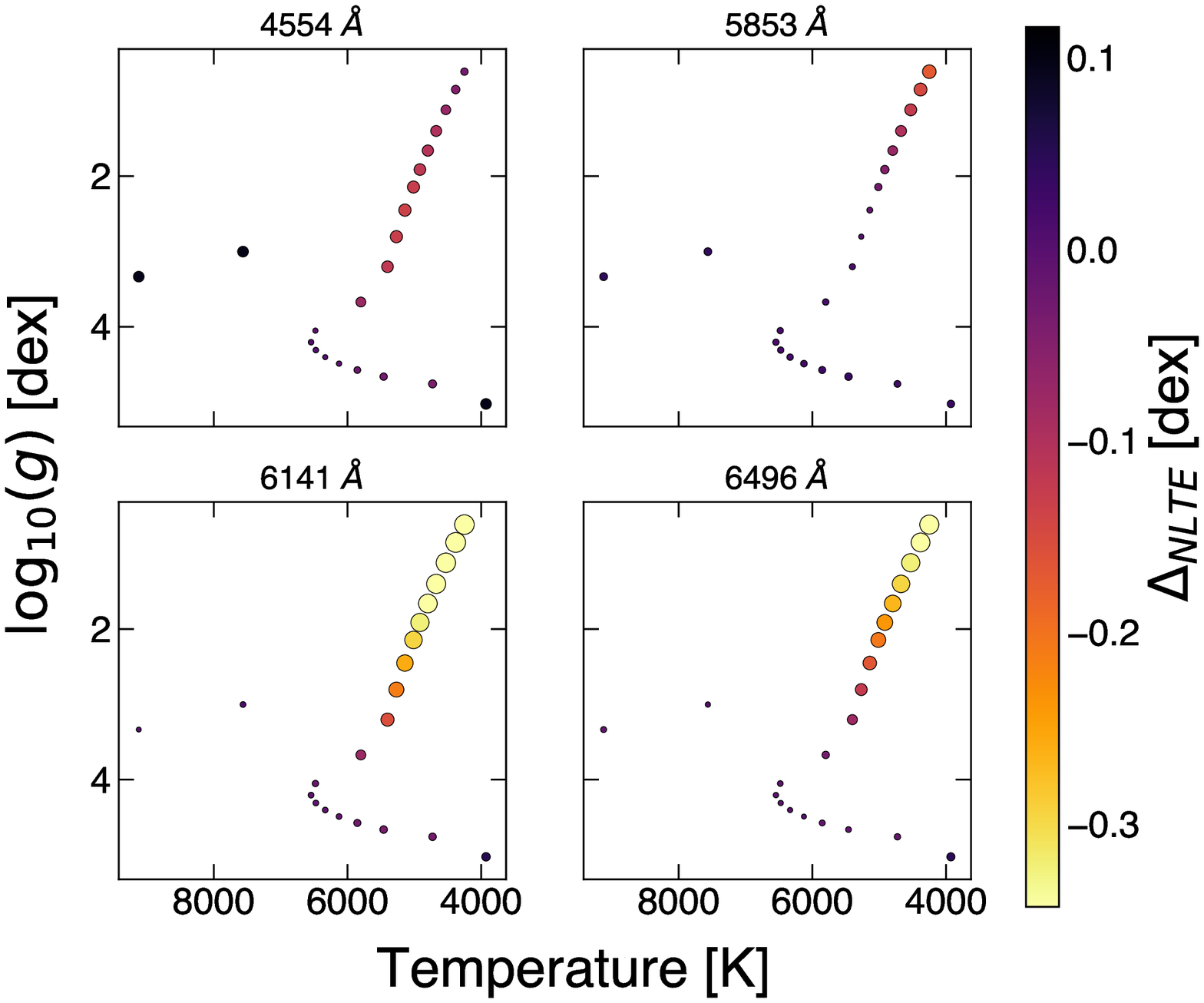}
   \caption[center]{barium}
              \label{f_HRD_Ba}%
\end{subfigure}
%
%
%
%
% Figure: HRD
%
\begin{subfigure}{.5\textwidth}
   \includegraphics[width=0.99\textwidth ,center]{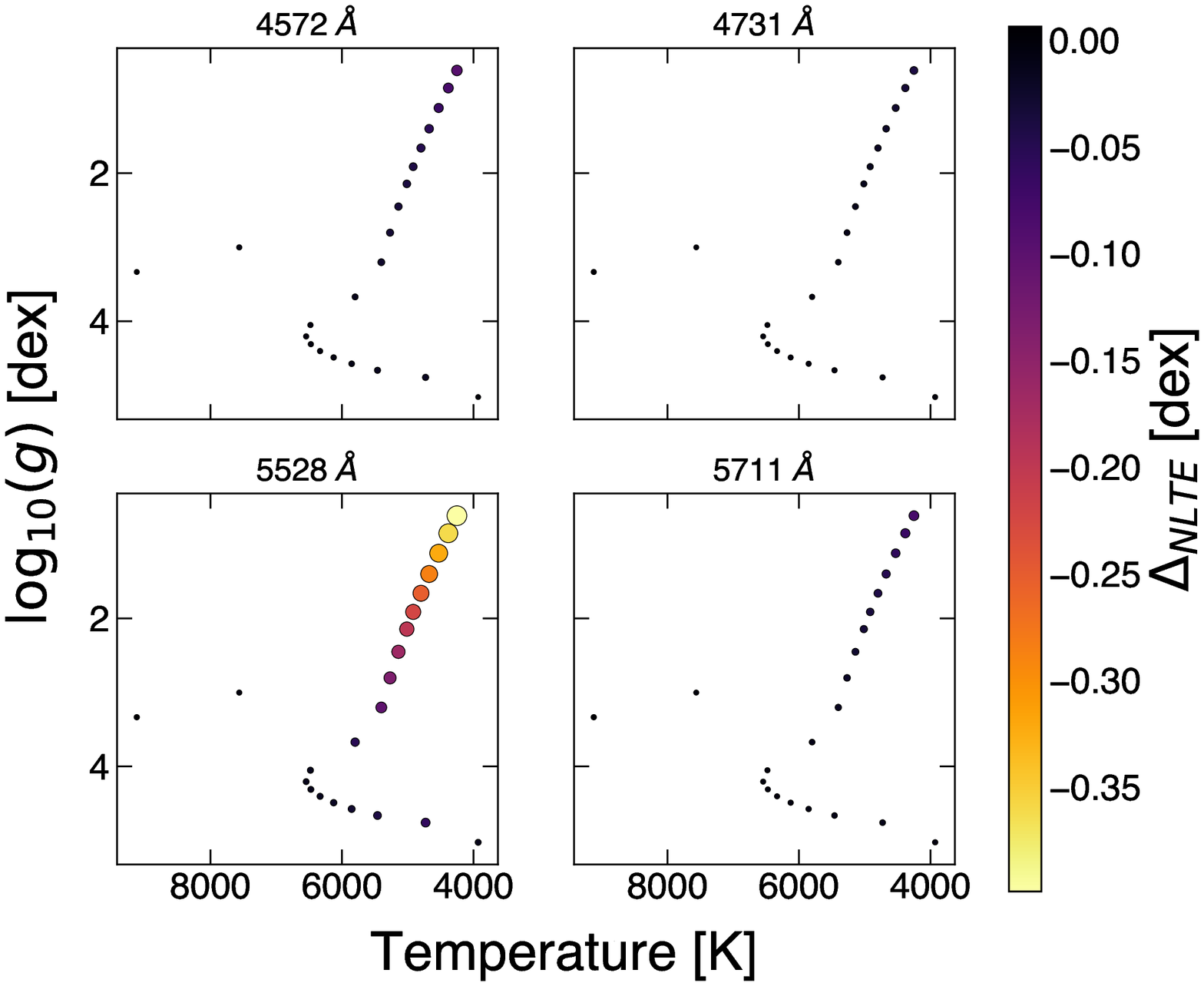}
   \caption[center]{magnesium}
              \label{f_HRD_Mg}%
\end{subfigure}
%
%
%
%
% Figure: HRD
%
\begin{subfigure}{\textwidth}
   \includegraphics[width=.5\textwidth ,center]{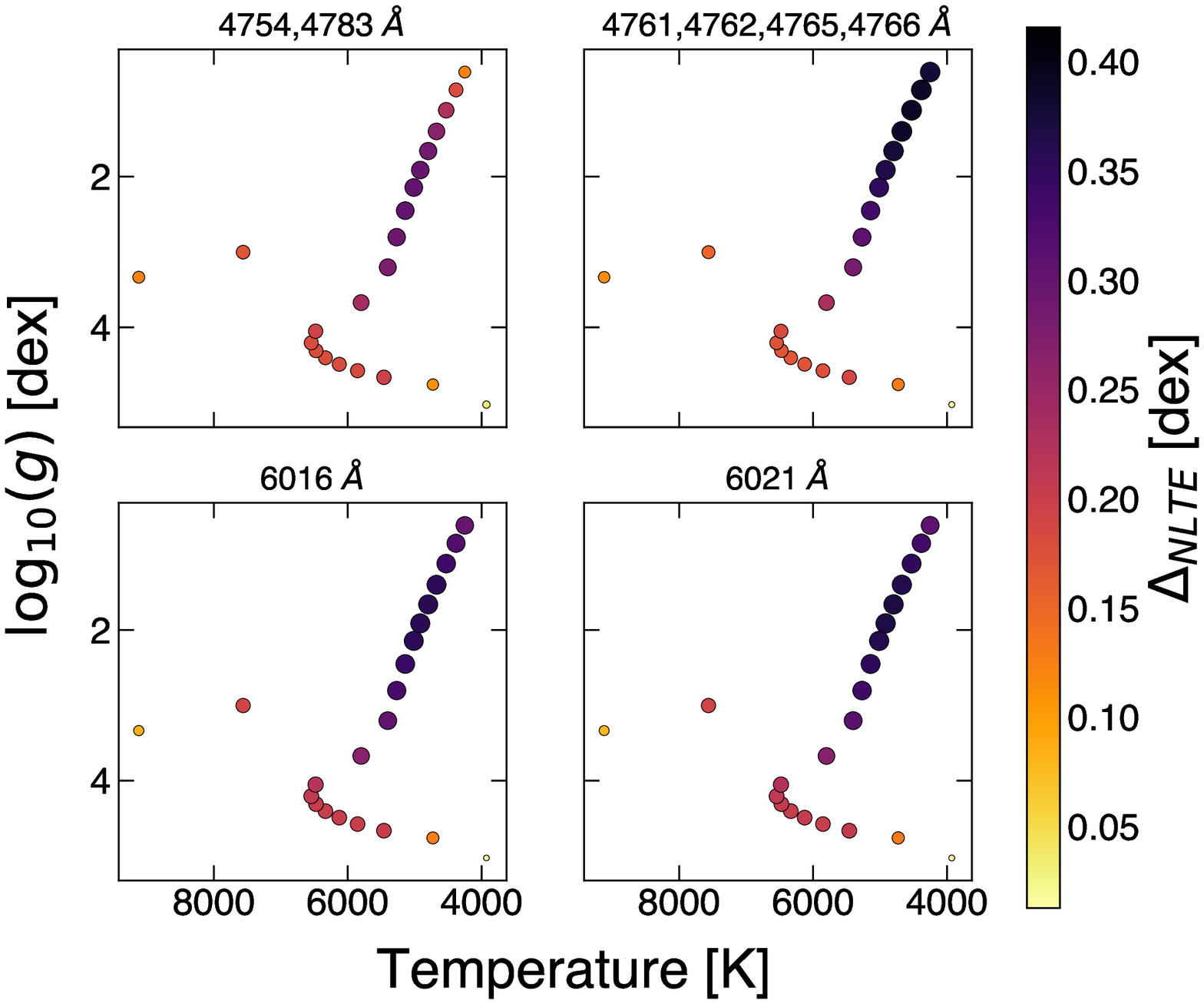}
   \caption[center]{manganese}
              \label{f_HRD_Mn}%
\end{subfigure}
\caption{Ba, Mg, and Mn NLTE abundance corrections in dex for all 22 model atmospheres in Table \ref{t_model_atmos} and transitions in Table \ref{t_lines}. The abundance correction is coded in colour and marker size. For Mn the lines are grouped as multiplets, the average correction per multiplet is shown here. The LTE reference is chosen to be a common measured abundance in GC's, that is $[Ba/Fe]=-0.2$, $[Mn/Fe]=-0.4$, $[Mg/Fe]=+0.4$. }
\label{f_HRD}
\end{figure*}

The plots in this appendix show the same data as discussed in Sect.\ref{ss_results_indiv_star}. In Fig.\ref{f_HRD} NLTE corrections are visualised as a function of surface gravity and effective temperature, arranged like a Hertzsprung-Russel diagram.
\clearpage
\section{Integrated light CoG}
\begin{figure*}[!hb]
\begin{subfigure}{.5\textwidth}
   \includegraphics[width=0.99\textwidth,center]{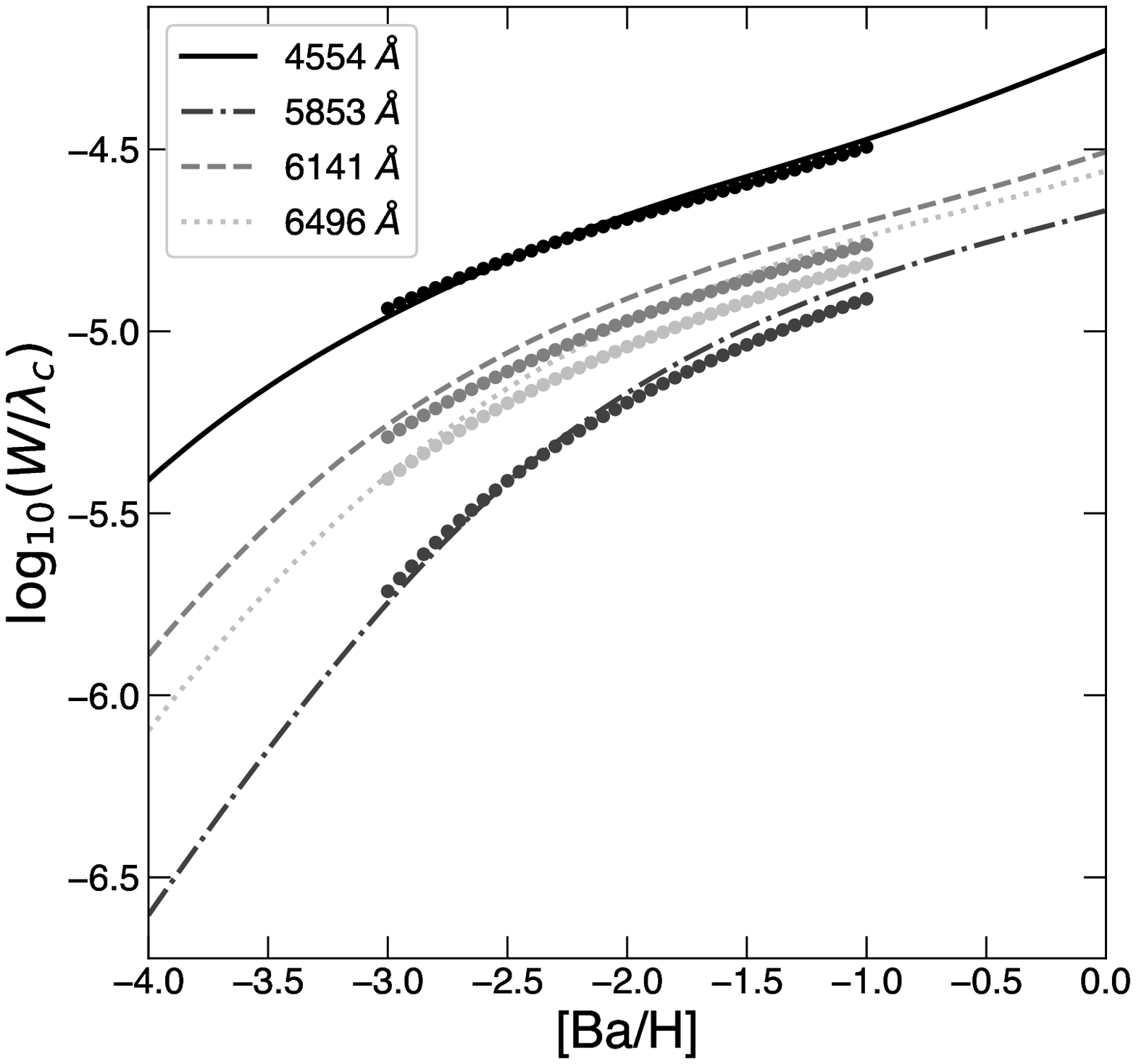}
   \caption[center]{barium}
              \label{f_COG_Ba}%
\end{subfigure}
\begin{subfigure}{.5\textwidth}
   \includegraphics[width=0.99\textwidth,center]{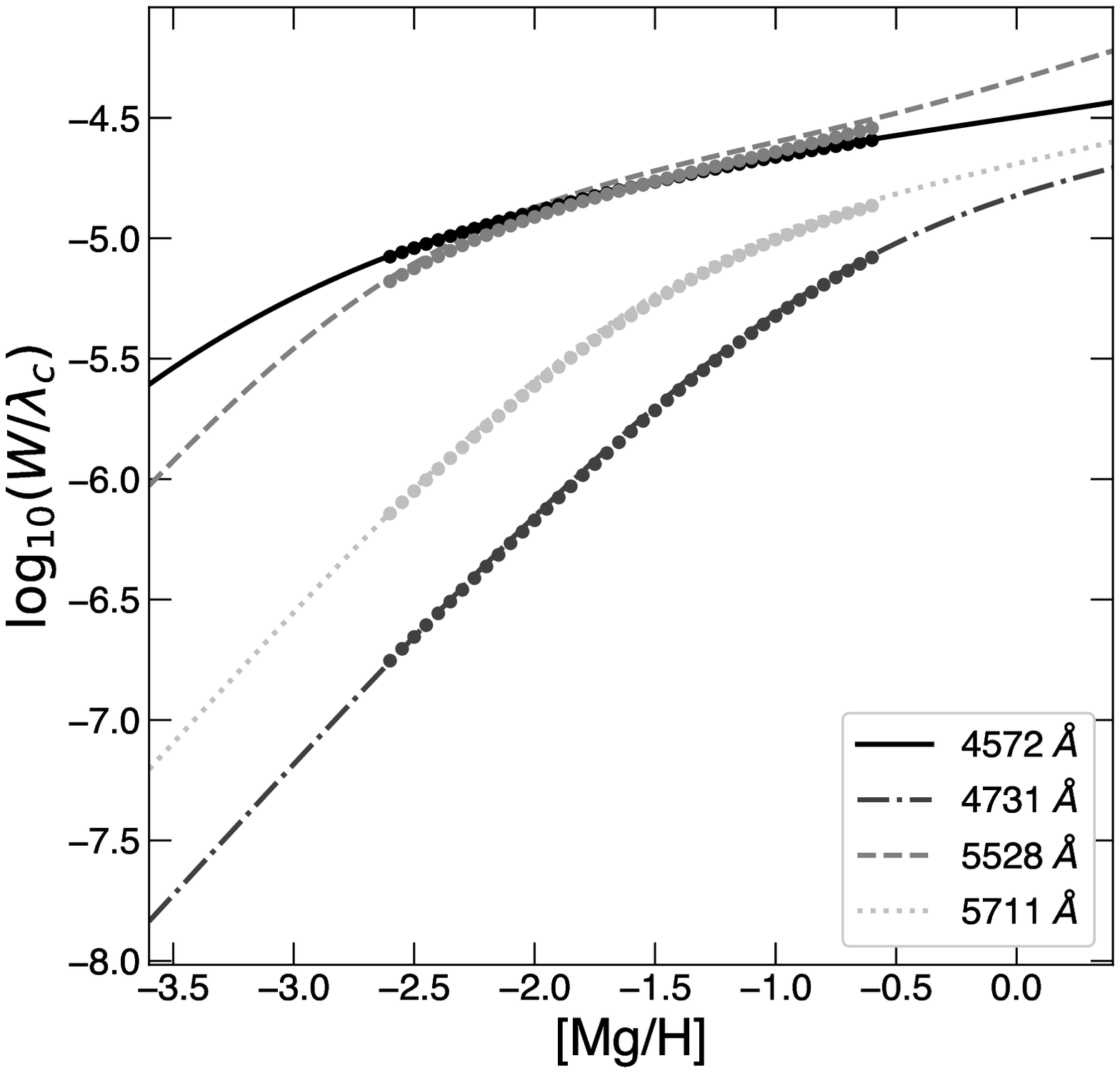}
   \caption[center]{magnesium}
              \label{f_COG_Mg}%
\end{subfigure}
\begin{subfigure}{\textwidth}
   \includegraphics[width=0.5\textwidth,center]{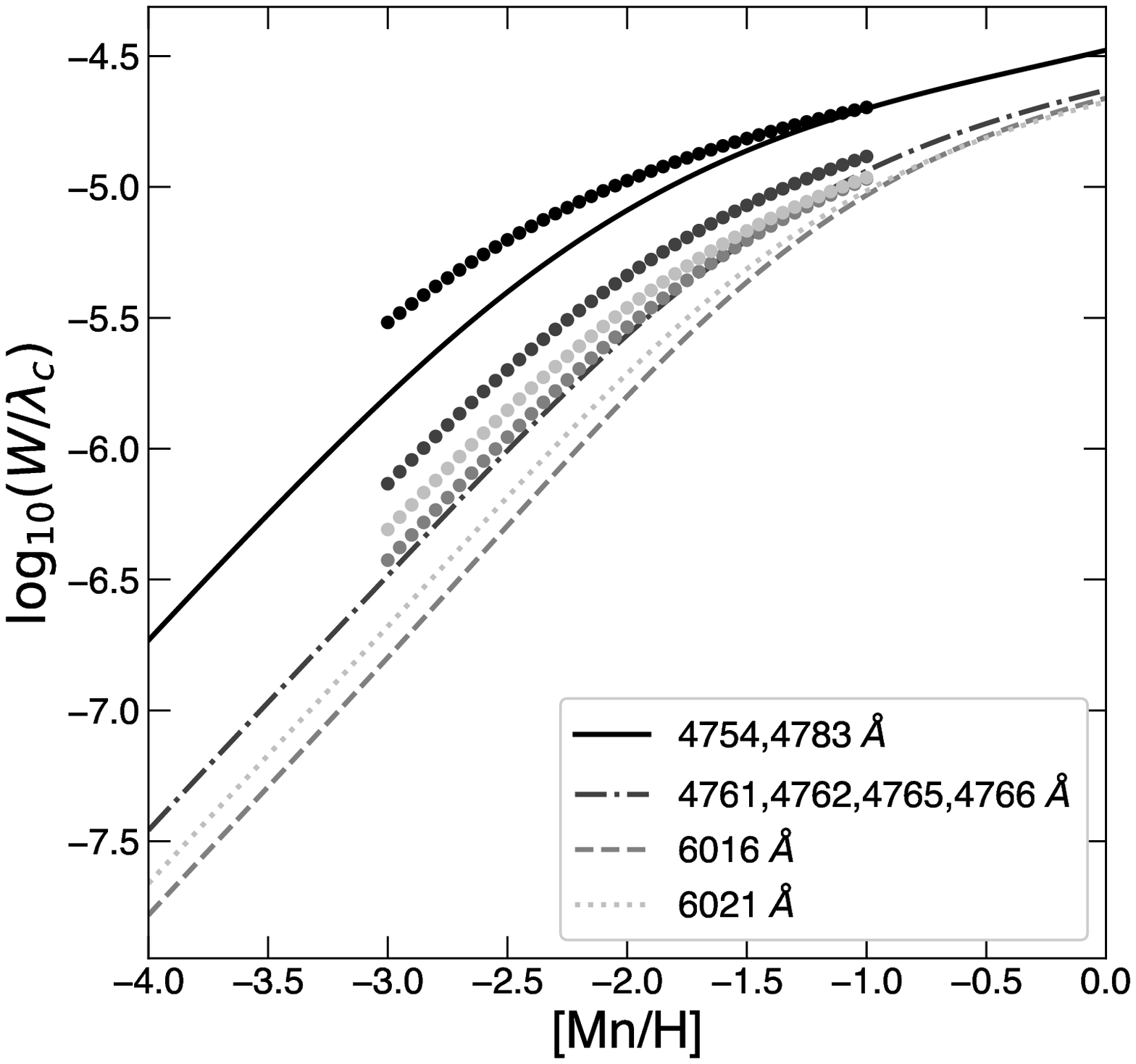}
   \caption[center]{manganese}
              \label{f_COG_Mn}%
\end{subfigure}

\caption{Ba, Mg, and Mn curve of growth for the integrated light spectra. For Mn the average curve per multiplet is shown. The dots refer to the LTE reference equivalent width-abundance combination. In every sub-plot $\lambda_c$ refers to the central wavelength of each curve, which is the transitions in Table \ref{t_lines}.}
\label{f_COG}

\end{figure*}

As mentioned in Sect.\ref{ss_results_iteg_light} the CoGs arsing from the integrated light analysis can be found in this appendix, Fig.\ref{f_COG}. Equivalent to App.\ref{appendix_lte_ref} and Sec.\ref{ss_ref_depend}, also in the integrated light the same behaviour of the CoGs as seen in Fig.\ref{f_cog_ref_comp} can be found.

%
% Figure: COG integrated Light
%
 
\end{document}